\documentclass[tightenlines,twocolumn,eqsecnum,floats,aps,amsmath,amssymb,nofootinbib,prl,shownopacs,floatfix,notitlepage]{revtex4-1}
\usepackage{graphicx, wrapfig}
\usepackage{amssymb}
\usepackage{color}
\usepackage{mathrsfs}
\usepackage{subfigure}
\usepackage{chngcntr}
\counterwithout{equation}{section}

\def\f{\frac}

\def\mpl{m_{\rm Pl}}

\def\rhopl{\rho_{\rm Pl}}

\def\rhomax{\rho_{_{\rm max}}}

\def\phib{\phi_{_{\rm B}}}
\def\phidb{\dot{\phi}_{_{\rm B}}}

\def\ig{\includegraphics}

\def\lLQC{\lambda_{\rm LQC}}

\usepackage{enumerate}

\newcommand{\be}{\nopagebreak[3]\begin{equation}}
\newcommand{\ee}{\end{equation}}
\newcommand{\bfig}{\nopagebreak[3]\begin{figure}}
\newcommand{\efig}{\end{figure}}
\newcommand{\ba}{\nopagebreak[3]\begin{eqnarray}}
\newcommand{\ea}{\end{eqnarray}}

\newcommand{\bmult}{\nopagebreak[3]\begin{multline}}
\newcommand{\emult}{\end{multline}}
\newcommand{\fref}[1]{Fig.\,\ref{#1}}
\newcommand{\eref}[1]{eq.\,(\ref{#1})}

\begin{document}
\preprint{IGC-15/9-3}
\title{Inflation with the Starobinsky potential in Loop Quantum Cosmology}
\author{B\'eatrice Bonga}
\email{bpb165@psu.edu }
\author{Brajesh Gupt}
\email{bgupt@gravity.psu.edu}

\affiliation{
Institute for Gravitation and the Cosmos \& Physics Department, The Pennsylvania State University, University Park, PA 16802 U.S.A.
}

\pacs{}
\begin{abstract}
A self-consistent pre-inflationary extension of the inflationary scenario with 
the Starobinsky potential, favored by Planck data, is studied using 
techniques from loop quantum cosmology (LQC). The results are compared with the
quadratic potential previously studied. Planck scale completion of the
inflationary paradigm and observable signatures of LQC are found to be robust
under the change of the inflaton potential. The entire evolution, from the 
quantum bounce all the way to the end of inflation, is compatible with 
observations. Occurrence of desired slow-roll phase is almost inevitable and 
natural initial conditions exist for both the background and 
perturbations for which the resulting power spectrum agrees with recent 
observations. There exist initial data for which the quantum 
gravitational corrections to the power spectrum are potentially observable. 
\end{abstract}

\maketitle

The inflationary scenario is highly successful in explaining, with minimal
assumptions, the primordial origin of structure formation and small 
inhomogeneities observed in the cosmic microwave background (CMB). There is a
plethora of inflationary models whose predictions can be matched with
observations. Recent data from Planck show that, for single field inflation, 
the Starobinsky potential is favored over others, such as the quadratic one 
\cite{planck15xx}. The standard inflationary models are based on classical 
general relativity and therefore inherit the big bang singularity. This leaves 
several conceptual issues unaddressed which are expected to be resolved by a 
fundamental quantum gravity theory. For instance: 
(i) Is there a consistent pre-inflationary extension that admits a
finite quantum gravity regime and confronts the problem of a past singularity? 
(ii) If so, does inflation occur naturally without the need of fine tuning the 
initial conditions? 
(iii) Do the perturbations remain in a Bunch-Davies state at the onset of
inflation?
(iv) How strongly do the predictions of such a fundamental theory depend on the
model of inflation?

Loop quantum cosmology (LQC) provides an excellent avenue to address these
issues. The underlying quantum geometric effects of LQC modify the Planck 
scale physics leading to the resolution of classical singularities in a variety 
of homogeneous and inhomogeneous cosmological settings \cite{aps1,aps3}
(see, e.g., \cite{as1} for a review). 
In all these models, the big bang is replaced by a quantum 
bounce, evolution is deterministic and all curvature scalars remain finite 
throughout the evolution.\footnote{For other approaches to 
bouncing models see, e.g.
\cite{Khoury:2001wf,Gasperini:2002bn,Khoury:2001bz,Kallosh:2001du}. Also see
\cite{Battefeld:2014uga} (and references therein) for a recent review.} In this 
Letter, we investigate the aforementioned issues for single field inflation 
with the Starobinsky potential in LQC using the framework of \cite{akl,aan2} 
and compare our results with the quadratic potential studied in \cite{asloan_prob2,ck1,aan3}. 

Note that a priori there is no reason to believe that the pre-inflationary 
dynamics of the Starobinsky potential and quadratic potential are similar
because: 
(i) already their inflationary dynamics is quite different; and
(ii) the evolution equation for the scalar perturbations depends explicitly 
on the potential. 
We find that despite these differences, not only is the occurrence of slow-roll 
almost inevitable as is the case for the quadratic potential, but also the 
observable LQC corrections to the power spectrum remain similar to those for 
the quadratic potential. Hence, LQC predictions are robust. Furthermore, there are 
interesting signatures of the quantum geometry on long wavelength modes which can 
modify inflationary tensor fossils and can couple with observable modes as suggested in 
\cite{schmidthui,agulloassym} leaving imprints on the primordial power 
spectrum. This potentially opens new avenues to explore the origin of CMB 
anomalies in quantum gravity.

{\bf Framework:} In LQC, the background quantum geometry is described by a
wavefunction which is a solution of the quantum Hamiltonian constraint. For
sharply peaked wavefunctions, which are peaked at classical solutions at late 
times, the leading quantum corrections to the background spacetime can be 
captured by the effective description of LQC, given by the following 
modified Friedmann and Raychaudhuri equations \cite{vt,ps09,ag1}:
\ba
\label{eq:fried}
H^2 &:=& \left(\f{\dot a}{a}\right)^2 = \f{8\pi G}{3} \rho
\left(1-\f{\rho}{\rhomax}\right), \\ 
\label{eq:ray}
\dot H &=& -4\pi G \left(\rho + P\right) \left(1- 2\f{\rho}{\rhomax}\right),
\ea
where $H$ is the Hubble rate, $a$ is the scale factor,
$\rhomax=18\pi/\Delta_o^3~\rhopl\approx0.41\, \rhopl$ is the universal maximum
of the energy density and $\Delta_o \approx 5.17$ is the minimum eigenvalue of the area 
operator, whose value is fixed via black hole entropy calculations in
loop quantum gravity \cite{gamma1,gamma2}. The LQC evolution of spacetime is completely described by 
(\ref{eq:fried}) and (\ref{eq:ray}) which, given a matter source with the energy
density $\rho$ and pressure $P$ form a well posed initial value problem. 
Note that the LQC modifications are dominant only in the quantum gravity regime,
where the energy density of the matter field is Planckian. 
When the spacetime curvature is sub-Planckian ($\rho\ll\rhomax$), equations
(\ref{eq:fried}) and (\ref{eq:ray}) reduce to the classical Friedmann and
Raychaudhuri equations. 

In the inflationary model under consideration, the matter source is a scalar 
field with a self-interacting potential, which also drives inflation \cite{Barrow:1988xi,Barrow:1988xh,Maeda:1988ab,Starobinsky:2001xq,DeFelice:2010aj}:
\be
        \label{eq:pot}
        V(\phi) = \f{3M^2}{32 \pi G} \left(1- e^{-\sqrt{\f{16\pi G}{3}} \phi}
\right)^2,
\ee
where $M$ is the mass of the inflaton $\phi$ with
$\rho=\dot\phi^2/2+V(\phi)$ and $P=\dot\phi^2/2-V(\phi)$. This potential
is commonly known as the Starobinsky potential and is related to 
various other potentials, e.g. Higgs inflation, via $\alpha$-attractors \cite{Galante:2014ifa}. 
It is evident from the above expression that $V(\phi)$ increases exponentially for $\phi<0$, 
has a minimum at $\phi=0$ and approaches a constant value ($3M^2/32\pi G$) as 
$\phi\rightarrow\infty$. The form of the Klein-Gordon equation in LQC remains unchanged: 
\be
 \ddot \phi + 3 H \dot\phi + \partial_\phi V(\phi) = 0,
\ee
which reflects the fact that LQC corrections are purely quantum geometrical and
do not require any modification or violation of the standard energy conditions.

In the standard inflationary scenario, by using Einstein's equations 
and the slow-roll conditions together with data observed by the Planck 
mission \cite{planck15xx}, we are led to set $M=2.51 \times 10^{-6} \, \mpl$. 
In particular, we used the value of the amplitude of the scalar power spectrum, 
its running, and the constraints on the number of e-folds ($N_*$) from the time 
the reference mode $k_*$ exited the horizon to the end of inflation.\footnote{Here, in order to compute the mass parameter, we have assumed 
that the LQC corrections to the power spectrum are extremely small at the pivot 
scale $k_*$. This assumption does indeed hold in the numerical results discussed here. However, in principle, this is inconsistent and a proper way to address 
this would require significant numerical work along the lines of \cite{Agullo:2015tca}.
There the authors find that, while this assumption is conceptually important, it 
leaves the main results practically unchanged for the quadratic potential; 
we expect the same to be true for the Starobinsky potential.}
This procedure also defines the phase of {\it desired slow-roll} which is
required for compatibility with observations. 
Thus, this phase always refers to the last $N_*$ e-folds of inflation,
but of course a phase of accelerated expansion can precede the desired slow-roll
period. At the onset of the desired slow-roll, $\phi=1.08~\mpl$ and the first
two Hubble slow-roll parameters are $\epsilon=1.98 \times 10^{-4}$ and
$\delta=-1.73 \times 10^{-2}$. Inflation ends when $\epsilon=1$ which corresponds 
to $\phi=0.19 \, \mpl$.

To understand the evolution of quantum perturbations one needs a
formalism of quantum fields on {\it quantum} cosmological spacetime, which --at
first-- seems an intractable task. It turns out that within the test field 
approximation (that is, the backreaction of the perturbations on the background 
quantum geometry is negligible), surprising simplifications occur: The dynamics 
of the perturbations on the quantum FLRW geometry is completely equivalent to 
that of perturbations evolving on a `dressed' FLRW metric, where `dressing'
refers to quantum corrections \cite{akl,aan2} (for other approaches within LQC, see 
\cite{FernandezMendez:2012vi,Gomar:2015oea,Barrau:2013ula,Cailleteau:2011kr,Barrau:2014maa,Bojowald:2011aa}). Interestingly, for sharply 
peaked states, the dressed metric is extremely well described by 
\eref{eq:fried} and (\ref{eq:ray})  \cite{ag1,aag1}.

{\bf Quantum bounce and slow-roll:} The key feature of homogeneous and isotropic
spacetimes in LQC is the occurrence of a quantum bounce when the energy density 
of the matter field reaches its universal maximum, i.e. $\rho=\rhomax$. Here, 
the Hubble rate $H$ is zero (\eref{eq:fried}), $\dot H$ 
is positive (\eref{eq:ray}) and therefore the scale factor has a minimum.
Following the bounce, there is a phase of super-inflation during which the
energy density decreases while the Hubble rate increases very rapidly and 
takes its maximum value at $\rho=\rhomax/2$. In 
further evolution, both the energy density and the Hubble rate continue to 
decrease monotonically. A typical time line of the post bounce evolution is
shown in \fref{fig:timeline}.
Inflation starts when the energy density of the matter field is of the order of
$10^{-12}\,\rhopl$ where LQC effects are negligible (for a detailed analysis of
the background evolution starting from the bounce till the end of inflation, 
see the accompanying paper \cite{bg2}). 
Hence, LQC provides a Planck scale completion of the 
inflationary scenario while resolving the classical singularity. The question
now is: How natural is the occurrence of inflation in this model? That is, what
fraction of the initial data surface leads to the desired slow-roll.
\bfig
\ig[width=0.47\textwidth]{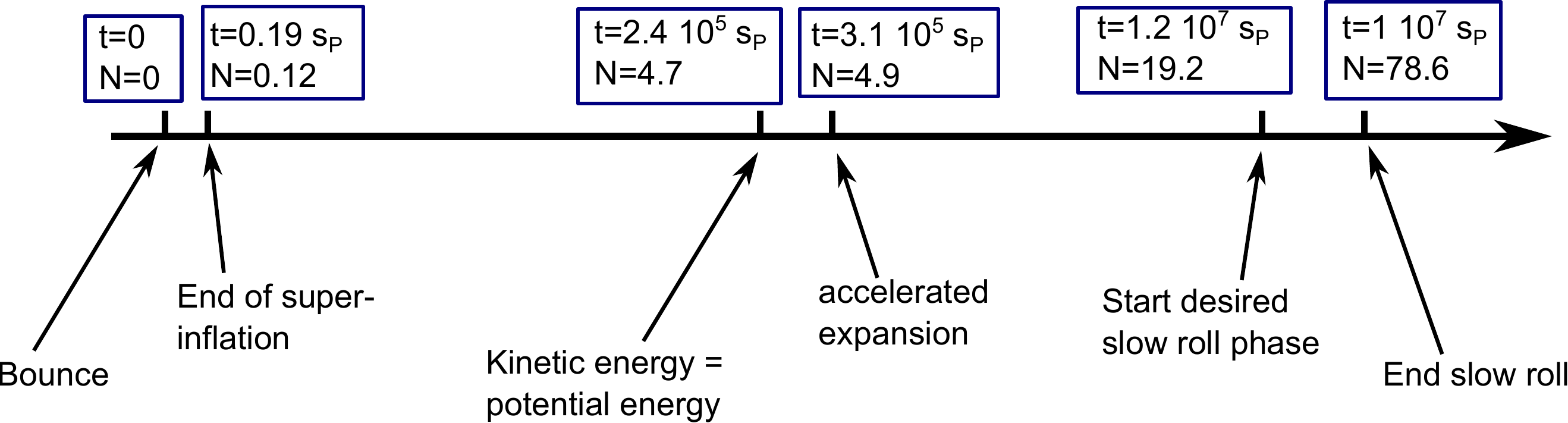}
\caption{Time line of evolution of the background spacetime with
$\phib=-1.41\,\mpl$ and $\phidb>0$.}
\label{fig:timeline}
\efig

To understand this, let us consider the space of initial conditions at the 
bounce which is 4 dimensional: $(a_{\rm B},\,H_{\rm B},\,\phib,\,\phidb)$. 
Utilizing the rescaling freedom of the scale factor we fix $a_{\rm B}=1$ without
loss of any generality. Furthermore, since $H_{\rm B}=0$, there are only two 
degrees of freedom in the space of initial conditions at the bounce: $\phib$ 
and $\phidb$. The energy density at the bounce is bounded by $\rhomax$ 
which yields: $\rhomax= \dot{\phib}^2/2+V(\phib)$. Hence, the entire space of 
initial conditions at the bounce is captured by $\phib$ and $\phidb$ whose range is given by: $\phib \in \, \left[-3.47 \, \mpl, \infty \right)$ and 
$|\phidb| \leq 0.905 \, \mpl^2$. Now the question is: What fraction of this
space of initial conditions give the desired slow-roll phase in the
future evolution? It turns out that the observationally compatible initial 
conditions leading to the desired slow-roll phase are: 
$\phib \geq -1.45 \, \mpl$ for $\phidb>0$ and $\phib \geq 
3.63 \, \mpl$ for $\phidb<0$. The remaining initial conditions either do not 
give inflation at all or do not have a sufficient number of e-folds required 
for compatibility with observations. Unlike the quadratic potential, 
the space of initial conditions --defined by the constant energy density surface at the bounce-- is unbounded, because the plateau-side of the potential continues
all the way to infinity and is well below $\rhomax$. Therefore, for a uniform 
measure the volume of the whole space of initial conditions for the Starobinsky 
potential is infinite and it is difficult to quantify the likelihood of inflation, 
as was done for the quadratic potential in \cite{asloan_prob1,asloan_prob2,ck1,Corichi:2013kua}.\footnote{This issue could be resolved for potentials, which are related to Starobinsky potential via $\alpha$-attractors, that have a plateau for a finite range in $\phi$. For such potentials the space of initial conditions at the bounce will 
be compact and a regular measure suffices to talk about probabilities. For instance, the Higgs potential satisfies this criterion as it has a plateau region in the center and exponential walls on both sides.} However, since only a tiny fraction of initial conditions do not lead to inflation it seems that, {\it qualitatively}, most of the initial conditions do lead to the desired slow-roll phase.

Interestingly, all of these initial conditions are kinetic energy dominated 
at the bounce.
The small fraction of initial data that does not lead to the desired slow-roll 
phase are either potential dominated or do not have enough initial kinetic 
energy. This is in striking contrast with the quadratic potential, where in 
fact all potential energy dominated bounces lead to the desired slow-roll phase 
and have a huge number of e-folds from the bounce till the end of inflation. 
However, it is the kinetic energy dominated initial conditions for
which LQC corrections to the primordial power spectrum are in the observable 
range for the quadratic potential. For the kinetic energy dominated 
initial conditions the inflaton behaves essentially as a massless scalar 
field near the bounce and consequently, the details of the potential are irrelevant
there. Therefore, while the observational
consequences of the {\it inflationary} dynamics for the two potentials are quite
different, the {\it pre}-inflationary dynamics is very similar.

Let us now consider the evolution of quantum perturbations, described 
by the Fourier transform of the gauge invariant Mukhanov-Sasaki scalar mode 
$q_k$ and two tensor modes $e_k$, on the 
quantum geometry. 

{\bf Primordial power spectrum:} 
The evolution equation for  $q_k$ and $e_k$, where $k$ is the comoving 
wavenumber, propagating on the quantum modified dressed geometry is given by 
\cite{aan2,aan3}:
\ba 
	\label{eq:scalar}
 	&q''_k (\tilde{\eta})& + 2 \f{\tilde{a}'}{\tilde{a}} q'_k(\tilde{\eta}) + \left(k^2 + 
 	\tilde{\mathcal{U}}(\tilde\eta) \right) q_k(\tilde{\eta})=0,\\
	\label{eq:tensor}
	&e''_k (\tilde{\eta})& + 2 \f{\tilde{a}'}{\tilde{a}} e'_k(\tilde{\eta}) + k^2 			
	e_k(\tilde{\eta})=0,
\ea
where the prime denotes the derivative with respect to the dressed conformal 
time $\tilde \eta$ and $\tilde{\mathcal U}= a^2 \left( 6\pi G
\f{\dot\phi^2}{\rho} \, V(\phi) -
2 \sqrt{\f{6\pi G \dot\phi^2}{\rho}} \,
	 \f{\partial V}{\partial \phi} + \f{\partial^2 V}{\partial \phi^2}
\right)$ is the dressed effective scalar 
potential  \cite{aan2,aan3}.

\bfig
\ig[width=0.45\textwidth]{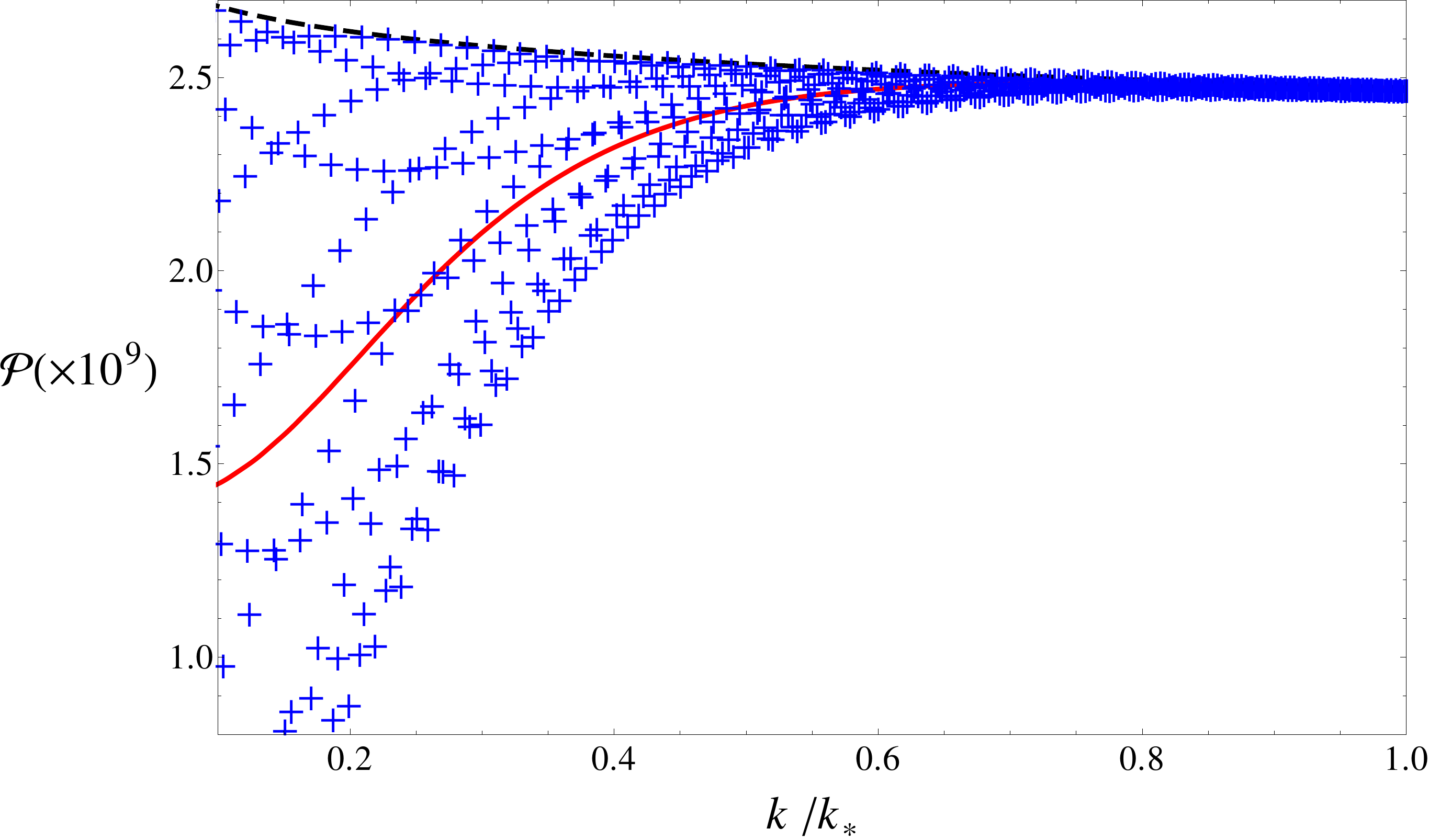}
\caption{Scalar power spectrum for Starobinsky potential with
$\phib=-1.42\,\mpl$ and $\phidb>0$. The blue `$+$' points show $\mathcal P_{\rm
LQC}$ which rapidly oscillates with $k$. The average is shown by (red) solid
curve, and $\mathcal P_{\rm BD}$ is shown by (black) dashed curve. 
}
\label{fig:power}
\efig

Recall that in the standard inflationary scenario the initial conditions for 
the quantum perturbations are given at the onset of inflation where the 
spacetime metric can be approximated by a de Sitter metric. Therefore, the 
modes of the perturbations are taken to be in the Bunch-Davies vacuum. In LQC,
on the other hand, since the pre-inflationary dynamics extends to the deep 
Planck regime where the spacetime metric cannot be approximated by a de Sitter 
metric, it is not meaningful to require that the modes be in the Bunch-Davies 
vacuum at the bounce. As described in \cite{aan2,aan3}, we choose
quantum perturbations to be in a 4th order adiabatic `vacuum' which has the
following two main properties: 
(i) respects the symmetry of the background spacetime and 
(ii) the expectation value of the renormalized energy density of the perturbations is
negligible with respect to the background energy density at the bounce. Unlike
in the de Sitter spacetime, this procedure does not single out a unique
state. 

A natural question then is: Among the admissible states {\it at the
bounce}, can one choose one that would lead to a power spectrum at the end of
inflation that not only agrees with the standard power spectrum at small 
angular scales but {\it also leads to a power suppression} at large angular 
scales, say, $\ell<30$?  \emph{Interestingly, the answer is in the affirmative} 
\cite{ag2}! This means that, when the state is evolved to the onset of  
slow roll, the $\ell <30$ modes are not in the Bunch-Davies vacuum at the onset of 
inflation.\footnote{Here, we only show existence of at least one 
state that leads to power suppression. It should be noted that there also exist 
states that show power enhancement for $\ell <30$. As of now, these states are at 
the same footing as the one chosen here that shows power suppression. 
The physical criteria to select states resulting in power suppression 
and the issue of their uniqueness are currently being investigated \cite{ag2}.}
The scalar power spectrum for such a vacuum state at the end of 
inflation is shown in \fref{fig:power} for initial background conditions 
$\phib=-1.42\,\mpl$ and $\phidb>0$. The deviation between LQC and standard power
spectrum can be understood by writing the LQC mode functions of the quantum 
perturbations at the onset of inflation as a Bogoliubov transformation on the 
Bunch-Davies states: 
$q(k) = \alpha(k)\,q_{\rm BD}(k)+\beta^*(k)\,q_{\rm BD}(k)$. The LQC power
spectrum can then be written as $\mathcal P_{\rm LQC}=(1+2 | \beta(k)|^2)
\mathcal P_{\rm BD}$, where $|\beta(k)|^2$ has the 
physical interpretation of the number density of the particles produced by the 
pre-inflationary dynamics with respect to the standard Bunch-Davies vacuum. 
For all choices of vacuum states, only modes with small $k$ deviate from the 
standard Bunch-Davies power spectrum, while for large $k$ there is remarkable 
agreement with the standard Bunch-Davies power spectrum. 
This is because for small $k$ the particle number 
density is non-zero ($|\beta(k)| > 0$) and $|\beta(k)|$ rapidly decays to zero 
for large $k$, as these modes are too energetic to be affected by 
background curvature and consequently do not get excited. 

This qualitative feature is also true for the quadratic potential 
and can be understood by 
comparing the relevant scales. During the evolution through the quantum gravity 
regime the quantum perturbations can interact with the curvature and become 
excited. 
Modes whose physical wavelength is smaller than the 
characteristic curvature length scale $\lLQC:=\sqrt{24\pi^2/\mathcal R_{\rm B}}$, 
where $\mathcal R_{\rm B}$ is the Ricci scalar at the bounce, propagate as if 
they are in flat space. Modes with physical wavelength larger than $\lLQC$ are 
affected by the background curvature: particle creation occurs and 
these modes become excited. Therefore, they no longer are in the Bunch-Davies 
vacuum state at the onset of inflation. Consequently, the primordial power 
spectrum for modes with small $k$ shows deviation from the power spectrum 
obtained in the standard inflationary scenario, where one assumes all modes to 
be in the Bunch-Davies state at the onset of inflation. Hence, the {\it origin 
of the non-Bunch Davies state is a direct consequence of the pre-inflationary
LQC dynamics} and occur always for small $k$ regardless of the details of 
the inflationary model. However, the quantitative details of the 
power spectrum do depend on the type of potential and the nature of the bounce. 
For instance, if the bounce is kinetic energy dominated, the scalar field behaves like a massless scalar field in the quantum gravity regime. In such cases, the LQC corrections to the power spectrum will be similar for all potentials. On the other hand, if the bounce is not kinetic energy dominated, then there could be potential specific signatures on the LQC corrections to the power spectrum. Thus, the LQC corrections to the power spectrum are robust under the change of potential for all kinetic energy dominated bounces including
the phenomenologically interesting initial conditions for the Starobinsky potential considered here and the quadratic potential.

{\bf Imprints on long wavelength modes:} So far in this Letter, we have seen that
the LQC corrections to the observable power-spectrum remain robust under the
change of potential. Nonetheless, there are interesting signatures of LQC 
pre-inflationary dynamics on super horizon modes for Starobinsky potential. 

From \eref{eq:scalar} and (\ref{eq:tensor}) it is immediately obvious that, 
just as in the standard inflationary paradigm, evolution of the scalar and 
tensor modes is different due to the presence of an effective potential 
$\tilde{\mathcal U}$ for the scalar evolution equation. Therefore, the particle 
density for scalar and tensor modes should be different from each other for 
$k^2 \lesssim \tilde {\mathcal U}$. It turns out that the numerical value of 
$\tilde {\mathcal U}$ is typically very small compared to even the smallest 
observable wavenumber $k_{\rm min}$ and consequently negligible. However, a small 
subset of the initial data surface ($-1.45~\mpl\lesssim\phib\lesssim-1.38~\mpl$)
exists for which $\tilde {\mathcal U}$ is of the order of $10^{-5} k_{\rm
min}^2$ so that for $k \lesssim k_{\rm min}/300$, the effect of $\tilde 
{\mathcal U}$ on the scalar modes is no longer negligible. The tensor and 
scalar particle densities are therefore different for these modes and 
$r_{\rm LQC} \neq r_{\rm BD}$, where $r$ is the tensor-to-scalar ratio and the 
subscript `BD' refers to the predictions from the standard inflationary 
paradigm. This is distinct from the quadratic potential, where this difference 
is negligible for all $k\gtrsim 10^{-5} k_{\rm min}$. This is shown in 
\fref{fig:rlqcbd}, where $r_{\rm LQC}/r_{\rm BD}$ is plotted for the 
Starobinsky and the quadratic potential.

\bfig
\ig[width=0.44\textwidth]{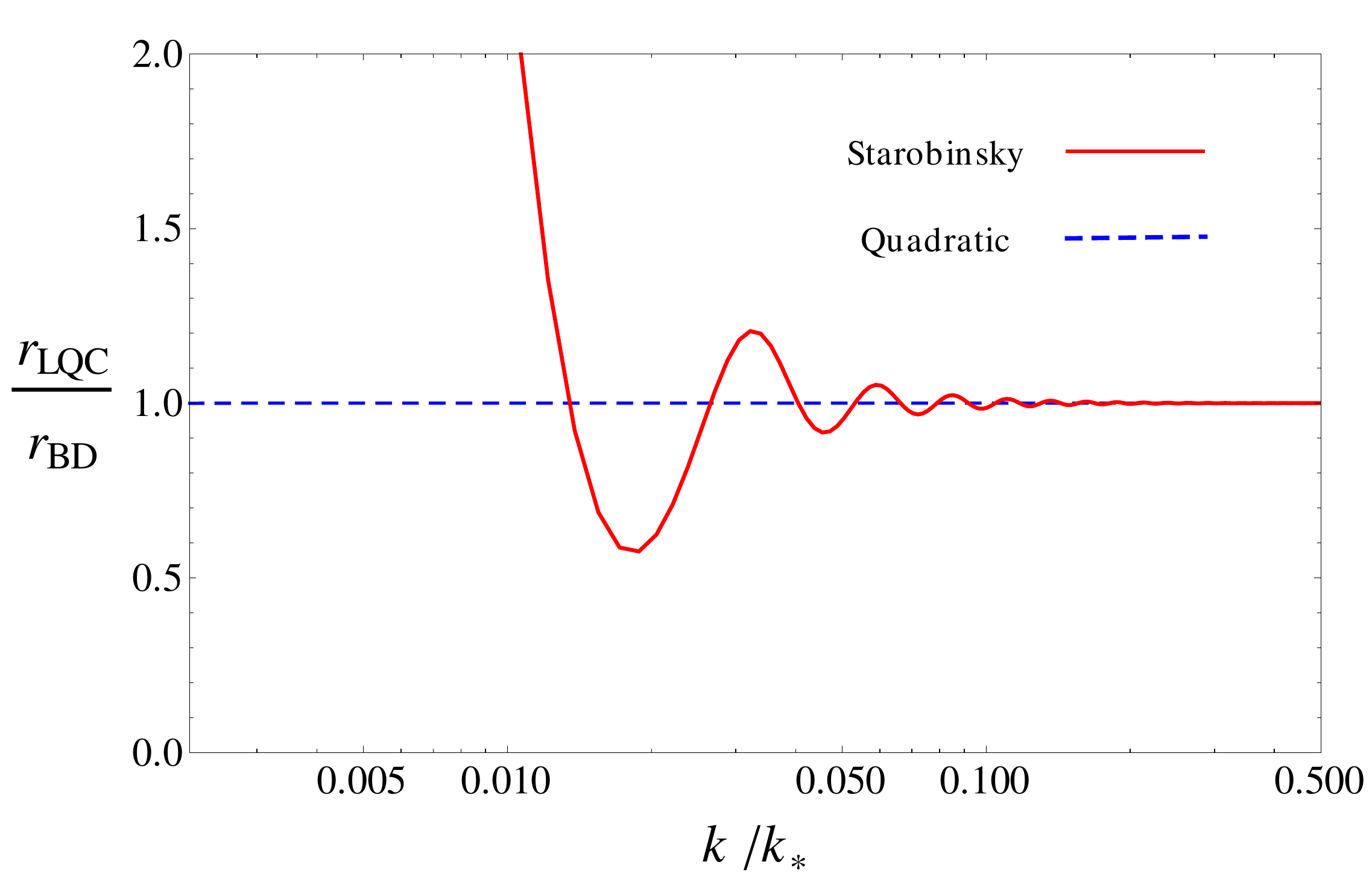}
\caption{Comparison of LQC corrections to tensor-to-scalar ratio between the 
Starobinsky and quadratic potential. States are chosen to give the maximum 
power suppression as in \fref{fig:power}.}
\label{fig:rlqcbd}
\efig

The quantum gravity induced deviations from $r_{\rm BD}$ 
are not directly relevant for observational modes, but the altered behavior of 
these super horizon modes can change the observed three-point functions (and 
higher order correlation functions) through mode-mode coupling as well as play 
an important role for tensor fossils. This difference between the scalar and 
tensor modes as compared to the standard picture, could thus lead to signatures 
in the non-Gaussian modulation of the power spectrum due to super horizon modes 
\cite{schmidthui,agulloassym}. It is noteworthy that the set of 
initial conditions for these effects is very small. Nonetheless, it is interesting as 
these initial conditions fall nicely into the regime for which the LQC corrections 
to the power spectrum are in the observable window. Thus, if these LQC corrections 
are observed, there is also hope to observe this effect. Furthermore, these initial 
conditions fall on the exponential side of the potential and not on the plateau side.
Therefore, they are very likely to be present for potentials which have exponential 
walls such as the Higgs potential.

In summary, we have studied a quantum gravitational extension of the
inflationary scenario with the Starobinsky potential using the framework of
quantum fields on quantum cosmological spacetime in LQC. Since the
pre-inflationary dynamics of LQC is very different from that of classical GR, it
is not obvious whether inflation is even obtained if one starts in the deep Planck
regime. If there is inflation, does it agree with observations? Are there any deviations 
from the standard power spectrum? We
found that the pre-inflationary dynamics of LQC fits very well with inflation 
and there are natural initial conditions for both the background spacetime and 
perturbations that lead to the desired slow-roll phase compatible with 
observations. Almost all initial conditions starting at the bounce lead to the
desired slow-roll phase. The LQC pre-inflationary excites the modes of the
perturbations, as a result they carry excitations over the Bunch-Davies state at 
the onset of inflation, giving corrections to the standard inflationary power
spectrum. There exist initial conditions for which these 
LQC induced corrections to the standard inflationary predictions at large 
angular scales are observable while being in complete agreement with 
observations at small angular scales. 

Since the observationally relevant initial conditions are all kinetic energy 
dominated in the quantum gravity regime, the inflaton behaves essentially as a 
massless scalar field and the details of the potential do
not affect the quantum gravity induced corrections to the observable power 
spectrum. Hence, {\it the occurrence of desired inflationary phase and 
corrections to the primordial power spectrum are robust features of 
LQC}. Thus, LQC has matured enough 
to confront recent observational data while providing a new quantum gravity 
window to understand various CMB anomalies. A detailed analysis of the 
results and extensive phenomenological investigations of the model is
presented in \cite{bg2}. 

Finally, the model considered here, i.e. a minimally coupled 
scalar with the Starobinsky potential is on-shell conformally related to a higher derivative 
gravitational theory whose Lagrangian density is $R + R^2/6M^2$ \cite{Barrow:1988xi,Barrow:1988xh,Maeda:1988ab,Starobinsky:2001xq,DeFelice:2010aj}. 
Here we studied the pre-inflationary 
dynamics of the scalar field with a standard Einstein-Hilbert action
where LQC quantization is well understood. An LQC treatment of the conformally related theory requires a careful study of loop quantization of higher
derivative theories, which is currently being pursued \cite{Zhang:2011vi}.

\acknowledgements{{\bf Acknowledgements:} We are grateful to Abhay Ashtekar for his constant guidance,
extensive discussions and feedback at various stages of the preparation of this 
manuscript, and to Ivan Agullo for extensive discussions. We would
also like to thank Aurelien Barrau and Parampreet Singh for 
fruitful discussions as well as the anonymous referees for their insightful comments which led to improvement in the manuscript.
This work was supported by NSF grant PHY-1505411, the 
Eberly research funds of Penn State and a Frymoyer Fellowship to BB.}
\vskip-0.6cm

\end{document}